\documentclass[pra,aps,twocolumn,superscriptaddress,showpacs]{revtex4}
\usepackage{graphicx}
\usepackage{amsmath}

\newcommand{\eq}{\begin{equation}}
\newcommand{\fine}{\end{equation}}

\begin{document}

\title{  An upper bound on the total inelastic cross-section as a function of
the total cross-section}

\author{Tai Tsun Wu }
\email{ttwu@seas.harvard.edu ,tai.tsun.wu@cern.ch } \affiliation{Harvard University, 
Cambridge, Massachusetts,and CERN,Geneva}

\author{Andr\'e Martin }
\email{martina@mail.cern.ch} \affiliation{Theoretical Physics Division,CERN, Geneva}

\author{Shasanka Mohan Roy}
\email{shasanka1@yahoo.co.in} \affiliation{Homi Bhabha Centre for Science Education, TIFR, V. N. Purav Marg, 
Mankhurd, Mumbai - 400 088.}

\author{Virendra Singh }
\email{vsingh@theory.tifr.res.in} \affiliation{Tata Institute of Fundamental Research, Mumbai 400005}

\begin{abstract}

Recently Andr\'e Martin has proved a rigorous upper bound on the inelastic cross-section $\sigma_{inel}$
at high energy which is one-fourth of the known Froissart-Martin-Lukaszuk upper bound on $\sigma_{tot}$. Here we 
obtain an upper bound on $\sigma_{inel}$ in terms of $\sigma_{tot}$ and show that the Martin bound on 
$\sigma_{inel}$ is improved significantly with this added information.

\end{abstract}

\pacs{03.67.-a, 03.65.Ud, 42.50.-p}

\maketitle

\section*{\large{1. Introduction}}
The total cross-section $\sigma_{tot} (s)$ for two particles to go to anything 
at c.m. energy $\sqrt s$ must obey the Froissart-Martin bound,
\begin{equation}
\sigma_{tot} (s) \leq_{s\rightarrow \infty} C \> [\ln (s/s_0)]^2
\end{equation}
proved at first from the Mandelstam representation by Froissart \cite{Froissart1961} 
and later from the basic principles of axiomatic field theory by Martin \cite{Martin1966}.
Of the two unknown constants the constant $C$ was fixed by \cite{Lukaszuk-Martin1967} to
obtain,
\begin{equation}
 \sigma_{tot} (s) \leq_{s\rightarrow \infty} 4\pi/t_0 \> [\ln (s/s_0)]^2,
\end{equation}
where, $t=t_0$ is the lowest singularity in the $t$-channel. For many physically interesting 
cases such as $\pi \pi ,KK, K\overline K,\pi K, \pi N, \pi \Lambda$ scattering 
$t_0 = 4 m_{\pi}^2 - \epsilon$, $\epsilon $ being an arbitrary 
small positive constant, and $m_{\pi} $ the pion-mass \cite{Bessis-Glaser1967}. In some cases we 
can take $\epsilon =0$ , e.g. for pion-pion scattering if the D-wave scattering length is 
finite \cite{Colangelo2000}. It will be convenient to denote the 
right-hand side of the bound on $\sigma_{tot (s) }$ as
\begin{equation}
 \sigma_{max} (s) = 4\pi/t_0 \> [\ln (s/s_0)]^2 .
\end{equation}
 
In equation (3) $s_0$ is unknown. However, if one assumes that the total 
and elastic cross-sections are increasing beyond a certain energy, or if 
one works with cross-sections averaged over a certain energy interval, one 
can, using fixed $t$ dispersion relations, fix the scale \cite{Martin2010a}. A reasonable guess is 
that $s_0$ lies between the square of the pion mass and the square of the 
nucleon mass. This means an uncertainty of $\pm$ 10\% at the present 
energy of the LHC.\\
The Froissart-Martin bound has been seminal both to the development of the field of high 
energy theorems in axiomatic field theory (see e.g. the review \cite{Roy1972})and to 
that of phenomenological models leading to accurate predictions of total and elastic 
cross sections before their experimental measurements \cite{Cheng-Wu1970}. 
Remarkably, one of us (A. M.) has recently obtained a 
bound on the total inelastic cross section at high energy \cite{Martin2009}, 
\begin{equation}
 \sigma_{inel} (s) \leq_{s\rightarrow \infty} \pi/t_0 \>[\ln (s/s_0)]^2 ,
\end{equation}
which is one-fourth of the bound $\sigma_{max} (s) $ on the total cross-section, thus improving the simple bound 
$\sigma_{inel} \leq \sigma_{tot} $. 

The present paper is inspired by Martin's bound on the inelastic cross-section.  
In fact T. T. Wu \cite {Wu2009} by extending Martin's variational calculation 
to incorporate a given total cross-section and independently S.M .Roy and Virendra Singh 
\cite {Roy2009}, by exploiting their previous upper bound on the 
differential cross section 
in terms of elastic cross-section, \cite {Singh-Roy1970},\cite {Roy1972a} realized that 
one could solve a more general problem: find a bound on the inelastic cross-section as a 
function of the value of the total cross-section. It is obvious that if the total cross section
vanishes the inelastic cross section also vanishes. but it is also
extremely plausible that if one maximizes the total cross section, the
important partial wave amplitudes will be imaginary and maximal so that,
from the unitarity condition, there is no room left for the inelastic
cross section which will receive only negligible contributions from the
tail of the partial wave distribution. 

The net result exhibiting both these features is the bound we present in this paper,

\begin{equation}
\label{inelastic}
 \Sigma_ {inel} (s) \leq_{s\rightarrow \infty} \Sigma_ {tot} (s) \bigl ( 1- \Sigma_ {tot} (s) \bigr ),
\end{equation}
where,
\begin{equation}
 \Sigma _{tot} (s) \equiv  \sigma_{tot} (s) / \sigma_{max} (s) ,
\end{equation}
and
\begin{equation}
 \Sigma _ {inel} (s) \equiv  \sigma_{inel} (s) / \sigma_{max} (s) .
\end{equation}
    
Maximizing wth respect to $\sigma_{tot} $ we get the factor 1/4 announced at
the beginning of this paper,i.e.
\begin{equation}
  \sigma_{inel} (s) \leq_{s\rightarrow \infty} \sigma_{max} (s) / 4 .
\end{equation}

     In Sec. 2 we summarise our notations and recall the basic results from axiomatic 
field theory. We then present two possible derivations of the 
bound on the inelastic cross-section in terms of total cross-section,
the direct variational approach in Sec. 3, and the approach using the 1970 
bound on the differential cross section in terms of the elastic cross-section 
\cite{Singh-Roy1970}, \cite{Roy1972a} in Sec. 4. Sec. 5 contains concluding remarks 
including directions for future work on high energy phenomenology.

\section*{\large{2. Basic Results from Axiomatic Field Theory }}

Let $F(s,t)$ be the elastic scattering amplitude for $ab \rightarrow ab$ at c.m. energy $\sqrt{s}$ 
and momentum transfer squared $t$ and be normalized such that the differential cross-section is 
given by
\begin{equation}
 \frac{d \sigma }{d \Omega } (s,t) = \bigl | \frac{F(s,t)} {\sqrt{s} } \bigr |^2
\end{equation}
 with $t$ being given in terms of the c.m. momentum $k$ and the scattering angle $\theta$ by the 
relation,
\begin{equation}
 t= -2 k^2 (1- \cos \theta ).
\end{equation}
Then, for fixed $s$ larger than the physical $s-$channel threshold, $F(s;\cos \theta ) \equiv F(s,t)$ is 
analytic in the complex $\cos \theta $ -plane inside the Lehmann-Martin ellipse \cite {Lehmann1958},
\cite {Martin1966}, with foci -1 and +1 and semi-major axis $\cos \theta _0 = 1+ t_0 /(2 k^2) $, where 
$t_0$ is independent of $s$. In fact, as mentioned already, $t_0 = 4 m_{\pi}^2 - \epsilon$ for many interesting cases.
Within the ellipse $F(s,t)$ has the partial wave expansion,
\begin{equation}
 F(s,t)=\frac{\sqrt s }{k} \sum_{l=0}^{\infty} (2l+1)a_l (s)P_l (1+t/(2k^2)),
\end{equation}
which converges absolutely and uniformly in $t$ for $ |t| < t_0 $ ; hence $F(s,t)$ is analytic in $t$ 
for $ |t| < t_0 $ . Unitarity implies that,
\begin{equation}
 Im a_l (s) \geq | a_l (s) |^2
\end{equation}
in the physical region. Further, \cite{Jin-Martin1964} for fixed $t$ in the region $|t|<t_0$ , 
$F(s,t)$ satisfies dispersion relations in $s$ with two subtractions. This implies, in particular, 
that the $s$-channel absorptive part for $0 \leq t < t_0 $ has the convergent partial wave expansion,
\begin{eqnarray}
 A(s,t) \equiv Im F(s,t) \nonumber \\
= \frac{\sqrt s }{k} \sum_{l=0}^{\infty} (2l+1) Im a_l (s)P_l (1+t/(2k^2)),
\end{eqnarray}
and obeys 
\begin{equation}
 \int _C^{\infty} ds A(s,t) /s^3 < \infty ,\> 0 \leq t < t_0 .
\end{equation}
 Hence, if we assume that $ A(s,t)$ is continuous in $ s$, there exist sequences of $s 
\rightarrow \infty $ such that
\begin{equation}
 A(s,t) < Const. \frac{s^2}{\ln (s/s_0)},\> 0\leq t < t_0 .
\end{equation}
For simplicity, in this paper, we deduce asymptotic bounds on $\sigma_{inel} (s)$ 
only for such sequences. Bounds on energy averages will be considered later to 
avoid this restriction.
\vspace{0.2cm}

\section*{\large{3. Variational Bound on Inelastic Cross-section in terms of Total Cross-section }}
Since $\sigma_{inel} = \sigma_{tot} - \sigma_{el} $, this problem is equivalent to finding a lower 
bound on $\sigma{el} $. Further,
\begin{eqnarray}
 \sigma_{el} (s) = \frac{4 \pi}{k^2} \sum _{l=0}^{\infty} (2l+1)|a_l(s)|^2 \nonumber \\
\geq \frac{4 \pi}{k^2} \sum _{l=0}^{\infty} (2l+1)(Im a_l(s))^2 \equiv \sigma_{el,im} (s).
\end{eqnarray}
So, it suffices to find a variational lower bound on $\sigma_{el,im} $. We vary the 
$ Im a_l(s) $ subject to the unitarity constraints 
\begin{equation}
 Im a_l(s) \geq \> 0 \>,
\end{equation}
to a given value of,
\begin{equation}
 \sigma _{tot} (s) =  \frac{4 \pi}{k^2} \sum _{l=0}^{\infty} (2l+1) Im a_l(s) \>,
\end{equation}
and to the constraint
\begin{eqnarray}
A(s,t_0) \equiv \frac{\sqrt s }{k} \sum_{l=0}^{\infty} (2l+1) Im a_l (s)P_l (1+t_0 /(2k^2)) \nonumber \\
<  Const. \frac{s^2}{\ln (s/s_0)} \>.
\end{eqnarray}
For simplicity, since we work at a fixed-$s$ , we suppress the $s$-dependence of $Im a_l (s)$,
$\sigma_{el,im} (s) $ and $ \sigma _{tot} (s) $ . Denoting ,
\begin{equation}
 z_0 = 1+ t_0 / (2 k^2),
\end{equation}
the lower bound on $ \sigma_{el,im} $ is obtained by choosing,
\begin{equation}
 Im a_l = \alpha \bigl ( 1-P_l (z_0) /P_{L+r} (z_0) \bigr ), for \> 0 \leq l \leq L \>,  
\end{equation}
and,
\begin{equation}
 Im a_l =0, for \> l > L \>,
\end{equation}
with the constants $0\leq r <1, \alpha >0 $ and the positive integer $L$ being fixed from the given 
value of $\sigma_{tot}$ and the given upper bound on $A(s,t_0)$. We omit the straight forward proof 
which is by direct subtraction of a $ \sigma_{el,im} $ with arbitrary partial waves obeying the 
given constraints from the variational result. After carrying out the summations over $l$, 
the constraint equations become,
\begin{eqnarray}
 \sigma_{tot} \>  k^2 /(4 \pi \alpha )=  \nonumber  \\
(L+1)^2  - (P_{L+1} ^{'}(z_0) + P_{L} ^{'} (z_0) )/P_{L+r} (z_0) \>,
\end{eqnarray}
and
\begin{eqnarray}
 A(s,t_0) \frac{k}{\alpha \sqrt{s}} = P_{L+1} ^{'}(z_0) + P_{L} ^{'} (z_0) - \nonumber \\
\frac {(L+1)^2 P_L ^2 (z_0) - (z_0 ^2 -1)(P_{L} ^{'} (z_0))^2  } {P_{L+r} (z_0) },
\end{eqnarray}
and the bound on $ \sigma_{el} $ becomes,
\begin{eqnarray}
 k^2 /(4 \pi \alpha ^2 ) \> \sigma_{el} \geq \> k^2 /(4 \pi \alpha ^2 ) \> \sigma_{el,im} \geq \nonumber \\
(L+1)^2  - 2 (P_{L+1} ^{'}(z_0) + P_{L} ^{'} (z_0) )/P_{L+r} (z_0)  + \nonumber \\ 
\frac {(L+1)^2 P_L ^2 (z_0) - (z_0 ^2 -1)(P_{L} ^{'} (z_0))^2  } {(P_{L+r} (z_0))^2 } .
\end{eqnarray}

At high energies, using $s/\sigma_{tot} \rightarrow \infty$, the two constraint 
equations yield easily that $L=O(\sqrt{s} \ln (s/s_0))$ ; we may therefore set $r=0$ and use the following 
approximations for the Legendre polynomials,
\begin{eqnarray}
\label{approx}
P_L (z_0)&=& I_0 (\xi) (1+O(L/s)),\xi \equiv (2L+1) \sqrt { (z_0 -1)/2 } \nonumber \\
P_{L} ^{'} (z_0)&=&(1/2) L \sqrt {s/t_0} I_1 (\xi) (1+O(L/s)) , \nonumber \\
I_{\nu }(\xi)&=&\frac {\exp {\xi} } {\sqrt{2 \pi \xi } }(1- (4 \nu^2 -1)/ (8 \xi) +...), 
\xi \rightarrow \infty , \nonumber \\
for & & s\rightarrow \infty,\> L/\sqrt{s} \rightarrow \infty ,\> L/s \rightarrow 0 \>,
\end{eqnarray}
where the $I_{\nu } (\xi)$ denote the modified Bessel functions. We then have,
\begin{eqnarray}
 \sigma_{tot} \>  k^2 /(4 \pi \alpha ) \approx \> L^2 -\frac{I_1(\xi)} {I_0(\xi) } L \sqrt{s/t_0} \\
\approx \>  L^2 (1+O(\sqrt{s}/L) ),
\end{eqnarray}
 \begin{eqnarray}
 A(s,t_0) \frac{k}{\alpha \sqrt{s}} \approx I_1(\xi)  L \sqrt{s/t_0} + L^2 \frac{I_1(\xi)^2 - I_0(\xi)^2}{I_0(\xi)} \\
\approx I_0 (\xi) \frac{L \sqrt{s} } {2\sqrt{t_0} } (1+O(\sqrt{s}/L) ).
 \end{eqnarray}
The asymptotic bounds on elastic and inelastic cross-sections become, with these approximations,
\begin{equation}
 k^2 /(4 \pi \alpha ^2 ) \> \sigma_{el} \geq \> L^2-2L \sqrt{s/t_0 }\frac{I_1(\xi)} {I_0(\xi) }+L^2 (1- 
(\frac{I_1(\xi)} {I_0(\xi) })^2 ),
\end{equation}
and 
\begin{eqnarray}
  k^2 /(4 \pi \alpha ) \> \sigma_{inel} \leq \> (1-2\alpha))(L^2-L \sqrt{s/t_0 }\frac{I_1(\xi)} {I_0(\xi) })+ 
\nonumber \\ 
\alpha L^2 (\frac{I_1(\xi)} {I_0(\xi) })^2 . 
\end{eqnarray}
We now use the assumed upper bound on $A(s,t_0)$ to evaluate $L, \alpha$ for high energies. We have,
\begin{equation}
 Const. s/(\sigma_{tot} \ln (s/s_0) ) = I_0 (\xi) \frac{\sqrt{s} } {2L\sqrt{t_0} } (1+O(\sqrt{s}/L) ),
\end{equation}
which yields ,
\begin{eqnarray}
\frac{L}{\sqrt{s}}= (1/(2 \sqrt{t_0})) \ln (\frac{s}{s_0 ^2 \sigma_{tot} }) (1+O(\ln (s/s_0))^{-1} )\\
 \alpha = \frac {\sigma_{tot}(s) } {\hat{\sigma}_{tot}(s) } (1+O(\ln (s/s_0))^{-1} )
\end{eqnarray}
where,
\begin{equation}
 \hat{\sigma}_{tot}(s) \equiv 4\pi/t_0 \> [\ln (\frac{s}{s_0 ^2 \sigma_{tot} })]^2 .
\end{equation}
Hence, we have the lower bound on elastic cross-sections,
\begin{equation}
\label{elastic}
 \sigma_{el} (s) \geq \frac{(\sigma_{tot}(s))^2 } {\hat{\sigma}_{tot}(s) }(1+O(\ln (s/s_0))^{-1} ). 
\end{equation}
Note that $\hat{\sigma}_{tot}(s)$ can be replaced by $\sigma_{max}(s)$ for $s\rightarrow \infty$,
except in the unrealistic case $\sigma_{tot}\rightarrow 0, for s\rightarrow \infty$ which leads to a small 
inelastic cross-section $\sigma_{inel}(s)\rightarrow 0 $. Hence, using  equation (\ref{elastic}),
the upper bound on the inelastic cross-section valid in all cases is ,
\begin{equation}
 \sigma_ {inel} (s) \leq_{s\rightarrow \infty} \sigma_ {tot} (s) \bigl ( 1- \Sigma_ {tot} (s) \bigr ),
\end{equation}
which leads to the announced bound on $\sigma_{inel}(s)/\sigma_{max}(s)$, given by equation (\ref{inelastic}).

\vspace{0.2cm}
\section*{\large{4. Upper Bound on Inelastic Cross-section from an Upper Bound on Differential 
Cross-section in terms of Elastic Cross-section }} 

We show here that the inelastic cross-section bound can also be derived as a corollary of 
an upper bound on the differential cross section in terms of the elastic cross-section, 
established by two of us \cite {Singh-Roy1970} many years ago,
\begin{equation}
 \frac{d \sigma }{dt } (s,t=0) \leq _{s\rightarrow \infty} \frac{\sigma_{el} (s) } {4t_0}
[\ln (\frac{s}{s_0 ^2 \sigma_{el} })]^2.
\end{equation}
This bound can also be written as \cite {Roy1972a},
 \begin{eqnarray}
 \sigma_{tot}[1+\bigl(\frac{Re F(s,t=0)}{Im F(s,t=0)}\bigr)^2]  \nonumber \\
 \leq _{s\rightarrow \infty} \frac{4\pi \sigma_{el}}{t_0 \sigma_{tot}} [\ln (\frac{s}{s_0 ^2 \sigma_{el} })]^2.
\end{eqnarray}
If the real part $Re F(s,t=0)$ is unknown we have the weaker bound,
 \begin{eqnarray}
 \sigma_{tot} &\leq _{s\rightarrow \infty}& \sqrt{\frac{4\pi}{t_0 }}\bigl[\sqrt{\sigma_{el}}
\bigl (\ln (\frac{s}{s_0 ^2 \sigma_{tot} })-\ln (\frac{\sigma{el}}{\sigma_{tot} })\bigr )\bigr ]\nonumber \\
&\leq _{s\rightarrow \infty}& \sqrt{\sigma_{el} \hat{\sigma}_{tot}} +(2/e)\sqrt{\frac{4\pi \sigma_{tot}}{t_0 }} \>,
\end{eqnarray}
where, in the last line we have used the elementary inequality,$\sqrt{x} \ln x \geq -2/e, for \> 0<x<1$. This equation 
yields a lower bound on $\sigma_{el}(s)$ for any asymptotic behaviour of $\sigma_{tot}(s)$. As noted in the 
last section,for deducing an upper bound on $\sigma_{inel}(s)$, it suffices to assume that $\sigma_{tot}$ does not 
vanish for ${s\rightarrow \infty}$. In particular, if $ \sigma_{tot} (s) > 16 \pi /(e^2 t_0)$, we have
  \begin{eqnarray}
 \sigma_{tot} (\sqrt{\sigma_{tot}}-(2/e)\sqrt{\frac{4\pi}{t_0 }}\>)^2 &\leq _{s\rightarrow \infty}& 
\sigma_{el} \hat{\sigma}_{tot} \nonumber \\
&\approx _{s\rightarrow \infty}& \sigma_{el} \sigma_{max} 
\end{eqnarray}
and hence the upper bound on the inelastic cross-section,
\begin{equation}
 \sigma_ {inel} \leq_{s\rightarrow \infty} \sigma_ {tot}\bigl [ 1- \Sigma_ {tot}
 (1- (2/e)\sqrt{\frac{4\pi }{t_0 \sigma_{tot}}}\>)^2 \bigr ],
\end{equation}  
which yields the desired bound (\ref{inelastic}) on the inelastic cross-section if 
$ \sigma_{tot}(s)\rightarrow \infty, for \> s \rightarrow \infty $.

\vspace{0.2cm}
\section*{\large{5. Conclusion}}
We have derived an asymptotic upper bound on the inelastic cross-section in terms of the 
total cross-section which improves Martin's recent bound \cite{Martin2009} when 
$ \sigma_{tot}(s) \sim C (\ln (s/s_0) )^2 $. Varying $\sigma_{tot}(s)$ over its allowed range 
we recover Martin's result $ \sigma_{inel} < \sigma_{max} /4 $ for some sequences of 
$ s \rightarrow \infty $ mentioned before.For applications to high energy phenomenology, it is desirable 
to remove the unknown scale factor $s_0$ in these bounds, as well as the restriction to special 
sequences of $ s \rightarrow \infty $. One way forward is to derive bounds on energy averages of 
$\sigma_{inel}(s)$ given energy averages of  $\sigma_{tot}(s)$ and $A(s,t_0)$. One of us now has 
definitive results on the analogous problem of finding bounds on energy averages of the 
inelastic cross-section, as well as of the total cross-section \cite {Martin2010b}.   



\vspace{0.2cm}

\begin{center}
\large{\bf Acknowledgements}
\end{center}

\vspace{1cm}

S.M.R. is Raja Ramanna Fellow of the Department of Atomic Energy , and V. S. is INSA Senior Scientist. S. M. R. and V. S. 
acknowledge support from the project \# 3404 of the Indo-French Centre for 
promotion of advanced research (IFCPAR/CEFIPRA); S.M.R., V. S. and T. 
T. W. thank Luis Alvarez Gaume for hospitality at CERN. We thank Tullio 
Basaglia for help in the submission and the revision of the manuscript.


\vspace{0.3cm}
\begin{thebibliography}{99}
\bibitem{Froissart1961} M. Froissart, Phys. Rev. {\bf 123}, 1053 (1961).
\bibitem{Martin1966} A. Martin, Nuov. Cimen. {\bf 42}, 930 (1966).
\bibitem{Lukaszuk-Martin1967} L. Lukaszuk and A. Martin, Nuov. Cimen. {\bf 52A}, 122 (1967).
\bibitem{Bessis-Glaser1967} J. D. Bessis and V. Glaser, Nuov. Cimen. {\bf 50}, 568 (1967).  
\bibitem{Colangelo2000} G. Colangelo, J. Gasser, and H. Leutwyler, Phys. Lett.{\bf B488},261 (2000).
\bibitem{Martin2010a} A. Martin, talk given at ITEP, Moscow, October 2010, and to be published.
\bibitem{Roy1972} S. M. Roy, Phys. Reports, {\bf 5C}, 125 (1972).
\bibitem{Cheng-Wu1970} H. Cheng and T. T. Wu, Phys. Rev. Letters {\bf 24},1456 (1970);
C. Bourrely, J. Soffer, and T. T. Wu, Phys. Rev. {\bf D19}, 3249 (1979) and Nucl. Phys. 
{\bf B247}, 15 (1984). See also, for instance, A. D. Kaidalov, L. A. 
Ponomarev, and K. A. Ter-Martirosyan, Sov. J. Nucl. Phys. {\bf 44}, 468 
(1986), and A.Donnachie, H.G. Dosch, P.V. Landshoff and O.Nachmann, 
Pomeron Physics and QCD, Cambridge University Press (2002).
\bibitem{Martin2009} A. Martin, Phys. Rev. {\bf D80}, 065013 (2009).
\bibitem{Wu2009} T. T. Wu, private communication to A. Martin, April 2009, and presentation 
at Martin's 80th birthday Fest, Aug. 27, 2009, CERN-GENEVA.
\bibitem{Roy2009} S. M. Roy, private communication to A. Martin, July 2009.
\bibitem{Singh-Roy1970} V. Singh and S. M. Roy, Ann. Phys. {\bf 57}, 461 (1970).
\bibitem{Roy1972a} S. M. Roy, Phys. Reports {\bf 5C}, 125 (1972), p.146, Eq. (4.6b). 
\bibitem{Lehmann1958} H. Lehmann, Nuovo Cimen. {\bf 10}, 579 (1958); Fortschr. Physik {\bf 6} 159 (1959).
\bibitem{Jin-Martin1964} Y. S. Jin and A. Martin, Phys. Rev. {\bf 135B}, 1375(1964).  
\bibitem{Martin2010b} A. Martin, in preparation.

\end{thebibliography}
\end{document}